\documentstyle[floats,aps,epsf]{revtex}

\tighten

\begin{document}
\draft
\title{Gravitational self force and gauge transformations}
\author{Leor Barack$^{1}$ and Amos Ori$^2$}
\address{
$^1$Albert-Einstein-Institut, Max-Planck-Institut f{\"u}r Gravitationsphysik,
Am M\"uhlenberg 1, D-14476 Golm, Germany.
$^2$Department of Physics, Technion---Israel Institute of Technology,
Haifa, 32000, Israel
}
\date{\today}
\maketitle

\begin{abstract}
We explore how the gravitational self force (or
``radiation reaction'' force), acting on a pointlike test particle in curved
spacetime, is modified in a gauge transformation. We derive the general
transformation law, describing the change in the self force in terms of the
infinitesimal displacement vector associated with the gauge transformation.
Based on this transformation law, we extend the regularization prescription
by Mino {\it et al.}\ and Quinn and Wald (originally formulated within the
harmonic gauge) to an arbitrary gauge. Then we extend the
method of mode-sum regularization (which provides a practical means for
calculating the regularized self force and was recently applied to the
harmonic-gauge gravitational self force) to an arbitrary gauge. We find that
the regularization parameters involved in this method are gauge-independent.
We also explore the gauge transformation of the self force from the
harmonic gauge to the Regge-Wheeler gauge and to the radiation gauge,
focusing attention on the regularity of these gauge transformations. We
conclude that the transformation of the self force to the Regge-Wheeler
gauge in Schwarzschild spacetime is regular for radial orbits and irregular
otherwise, whereas the transformation to the radiation gauge is irregular
for all orbits.
\end{abstract}

\pacs{04.25.-g, 04.30.Db, 04.70.Bw}




\section{Introduction}

\label{secI}

Recent works, by Mino, Sasaki, and Tanaka \cite{MST}, and by Quinn and Wald
\cite{QW} (MSTQW), established a formal framework for calculating the local
gravitational self force acting on a pointlike particle in curved spacetime.
In these works, a particle of small mass $m$ was considered, whose
gravitational field may be treated as a small perturbation to the (vacuum)
background metric. Such a finite-mass particle does not follow a geodesic of
the background geometry, as its interaction with its own gravitational field
gives rise to the exertion of a ``self force''. In the above works, a
general formal expression was obtained for the $O(m)$ self-force correction
to the geodesic equation of motion.

From the astrophysical point of view, the pointlike particle model and the
self-force phenomenon may be applicable to binary systems with an extreme
mass ratio. Of particular relevance are binary systems composed of a
solar-mass compact object orbiting a supermassive black hole (of the kind
now believed to reside in the cores of many galaxies). Such systems are
expected to serve as main targets for the proposed space-based gravitational
wave detector LISA (the Laser Interferometer Space Antenna), specializing in
the low frequency range below 1 Hz \cite{LISA}. Knowing the local self force
would be necessary, in general, for describing the orbital evolution in such
systems, and, eventually, for characterizing the consequent waveform of the
gravitational radiation emitted.

When considering a model of a pointlike particle, one unavoidably encounters
divergent quantities: the perturbed metric diverges at the location of the
particle, and the ``bare'' self force associated with the metric
perturbation turns out indefinite. One then has to deal with the fundamental
issue of {\em regularization}; namely, extracting the correct, physical self
force from the (indefinite) expression for the bare self force. The combined
works by MSTQW present three different physically-motivated methods of
regularization, all yielding the same formal expression for the physical
self force $F_{{\rm self}}^{\alpha }$. This expression can be written in the
schematic form\footnote{
Strictly speaking, both quantities on the right-hand side of Eq.\ (\ref
{Int-1}) are indefinite as they stand. In practice, one actually defines
these two quantities as vector fields in the neighborhood of the particle.
Then, the self force $F^{\alpha}_{{\rm self}}$ is obtained by taking the
(well defined and finite) limit of the difference $F^{\alpha{\rm {(bare)}}
}-F^{\alpha{\rm {(inst)}}}$ as the particle is approached. For simplicity,
we shall not use here this more strict formulation.}
\begin{equation}\label{Int-1}
F_{\rm self}^{\alpha}=F^{\alpha}_{\rm bare}-F^{\alpha}_{\rm inst}.
\end{equation}
Here $F^{\alpha}_{\rm bare}$ is the ``bare'' force, derived by applying
a certain differential operator [see Eq.\ (\ref{gg35}) below] to the full
metric perturbation produced by the particle, and $F^{\alpha}_{\rm inst}$
is the singular piece to be removed. According to MSTQW analyses, this
singular piece is to be constructed from the local, ``instantaneous'' part
of the metric perturbation in the harmonic gauge, i.e., the part directly
propagated along the light cone. The finite difference $F^{\alpha}_{\rm bare}
-F^{\alpha}_{\rm inst}$ represents the effect of the ``tail'' part of the
particle's gravitational perturbation---the part scattered off spacetime
curvature before interacting back with the particle. [The result by MSTQW is
formulated in terms of the retarded Green's function. The bare force is then
expressed as an integral (of a certain combination of Green's function
derivatives) along the entire worldline of the particle, while the
instantaneous part $F^{\alpha}_{\rm inst}$ arises from integration along
an infinitesimal, local piece of the worldline, that contains the momentary
particle's location.]

The first direct implementation of MSTQW's prescription for an actual
calculation of the self force was carried out recently by Pfenning and
Poission \cite{EricNew}, who considered the motion of a particle in a weakly
curved region of spacetime (Pfenning and Poission also calculated the
electromagnetic and scalar self forces acting on a particle endowed with
electric or scalar charges, respectively). To allow calculation of the
gravitational self force in strong field as well, Barack \cite{paperI}
recently introduced a method of multipole mode decomposition, based on the
formal result by MSTQW. This method of ``mode sum regularization'' was
previously developed \cite{Rapid} and tested \cite{MSRStest} for the toy
model of the scalar self force. We comment that a different mode-sum
approach to the gravitational self force was proposed by Lousto \cite{Lousto}
. 

The gravitational self force---unlike its electromagnetic or scalar
counterparts---is a gauge-dependent entity. This statement means that the
value of the self force is changed, in general, when the metric perturbation
to which it corresponds is being subject to a gauge transformation (i.e., an
infinitesimal coordinate transformation). If fact, the self force can be
nullified along any segment of the worldline by a suitable choice of the
gauge. Thus, any expression for the self force would be meaningless, unless
one is provided with the information about the gauge to which this force
corresponds. In MSTQW's analysis, the construction of the self force is
formulated {\em within the harmonic gauge}, and the resulting expression
(\ref{Int-1}) therefore describes the {\em harmonic gauge} self force.
Likewise, all implementations of MSTQW's analysis considered so far \cite
{EricNew,paperI} have been confined to the framework of the harmonic gauge,
and have yielded the harmonic gauge self force.

It is of great importance to understand the gauge dependence of the self
force and to figure out how to construct it in gauges other than the
harmonic: From the theoretical point of view, characterization of the
self-force's gauge dependence is essential for a better understanding of the
self force phenomenon; From the practical point of view, the harmonic gauge
is not the most convenient one for actual calculations, as in this gauge
perturbation theory has not been developed so far to the extent it has in
other gauges: In the Schwarzschild case, most analyses of metric
perturbations have been formulated so far within the Regge-Wheeler gauge
\cite{RW,Zerilli} (see, however, the recent mode decomposition of
Schwarzschild's metric perturbations in the harmonic gauge \cite{paperI}.)
In the Kerr case, so far the only practical approach for calculating the
(mode-decomposed) metric perturbations is Chrzanowski's method \cite
{Chrzanowski}, which is based on the radiation gauge.

The main purpose of this paper is to provide a general prescription for
calculating the gravitational self force in various gauges. To this end we
shall first construct the general transformation law describing the behavior
of the self force under a gauge transformation. Based on this transformation
law, we re-express MSTQW's result (\ref{Int-1}) in an arbitrary gauge. We
then re-formulate our method of mode sum regularization for a general gauge.

The transformation rule describing the gauge transformation of the self
force guarantees that the self force will be well-defined if (i) it was
regular in the original gauge, and (ii) the gauge transformation is
sufficiently regular (namely, the displacement vector $\xi ^{\mu }$ is
sufficiently regular at the particle's location). A priori there is no
guarantee that the transformation from the harmonic gauge to another desired
gauge will satisfy this regularity criterion. One of the objectives of this
paper is to explore the regularity of the self force in two commonly-used
gauges: the Regge-Wheeler gauge and the radiation gauge. We find that the
gauge transformations from the harmonic to these two gauges do not satisfy
the required regularity criterion. As a consequence, our general
transformation law does not yield a definite expression for the self force
in these two gauges (the exception is the situation of a radial orbit in a
Schwarzschild background, in which case the Regge-Wheeler self force is well
defined). We note that this irregularity of the gauge transformation has
been noticed independently by Mino\cite{Mino}.

This paper is arranged as follows. We start in Sec.\ \ref{secII} by
exploring the way the gravitational self force transforms under a general
gauge transformation. In Sec.\ \ref{secIII}, which is somewhat out of the
main course of our discussion, we consider the gauge transformation of
linear gravitational forces in general. We find that this transformation law
conforms with that of the gravitational self force. The general self-force
transformation law is then used in Sec.\ \ref{secIV} to generalize MSTQW's
expression for the regularized self force from the harmonic gauge to an
arbitrary gauge. We also re-formulate our method of mode sum regularization
for a general gauge. A few examples are provided in Sec.\ \ref{secV}, where
we consider the transformation of the self force from the harmonic gauge to
the Regge-Wheeler and to the radiation gauges. We find that in the
Schwarzschild background the Regge-Wheeler self force is well-defined for a
radial orbit, but is ill-defined for non-radial orbits. The situation with
the radiation gauge is even worse: It is ill-defined even for a static test
particle in flat space, and hence presumably also in all types of orbit in
Schwarzschild or Kerr spacetimes. Finally in Sec.\ \ref{secVI} we summarize
our main results and conclusions. We also discuss the indefiniteness of the
self force in the Regge-Wheeler and radiation gauges, and suggest
preliminary ways to overcome this difficulty.

Throughout this paper we use metric signature $(-+++)$ and geometrized
units $G=c=1$.


\section{Gauge transformation of the self force}

\label{secII}

Our first goal in this section is to clarify the origin of the gauge
dependence of the gravitational self force. Once this origin is well
understood, the derivation of transformation law for the self force becomes
rather straightforward.

In discussing the origin of the gauge dependence, we find it useful to take
the following point of view towards the gravitational self force kinematics:
A point-like particle moves on a background metric $g_{0}$ (e.g. the
Schwarzschild geometry), and we wish to describe the particle's orbit. The
particle, having a mass $m$, deforms the geometry, which is now described by
the new metric, $g=g_{0}+h$, where $h$ denotes the linearized metric
perturbation produced by the particle. We also know that generally the
particle will not follow a geodesic of $g_{0}$, due to its finite mass $m$.
Since no external force is assumed to be present, one might attempt the
simple point of view, according to which the particle moves on a ``geodesic
of the perturbed metric $g$''. This naive formulation, however, is
unsatisfactory (if not totally meaningless), because the perturbed metric $g$
is singular at the particle's location. We therefore must apply a different
framework for analyzing the particle's motion: Assume that on the perturbed
spacetime the particle follows a worldline $x^{\mu }(\lambda )$, where $
\lambda $ is an arbitrary monotonous parameter (we do not assume that $
\lambda $ is a proper time in $g$, because the latter is not defined, due to
the divergence of $h$). We now project the worldline $x^{\mu }(\lambda )$
onto the background metric $g_{0}$ on the basis of ``same coordinate
values'' [we presume here that a choice of a coordinate system has been made
in advance in each of the two spacetimes. Furthermore, we assume that the
coordinates in the two spacetimes are ``the same'' if the small perturbation
is ignored---which is equivalent to assuming that $h$ is small, i.e. $O(m)$].
The projection defines a worldline $x^{\mu }(\lambda )$ on the background
metric $g_{0}$, and we denote by $\tau $ the proper time along this
worldline (with respect to the metric $g_{0}$). This construction now
provides us with a natural definition of the self force: It is simply given
by the acceleration associated with the worldline $x^{\mu }(\tau )$ in $
g_{0} $, through Newton's second law:
\begin{equation}
F_{{\rm {self}}}^{\alpha }\equiv m\left( \frac{d^{2}x^{\alpha }}{d\tau ^{2}}
+\Gamma _{\mu \nu }^{\alpha }(x)\frac{dx^{\mu }}{d\tau }\frac{dx^{\nu }}{
d\tau }\right) .  \label{gg1}
\end{equation}
In this expression, the connection $\Gamma $ (just like the proper time $
\tau $) is taken with respect to the background metric $g_{0}$.

The origin of the gauge dependence of the self force is now obvious: Since $
g $ and $g_{0}$ represent different geometries, in principle there is no
unique way to project a point (or a worldline) from $g$ to $g_{0}$. In the
above formulation---as well as throughout this work---we adopt the rule of
``same coordinate values''. Suppose now that an infinitesimal gauge
transformation is carried out in the perturbed geometry $g$, associated with
an infinitesimal displacement vector $\xi ^{\mu }$:
\begin{equation}
x^{\mu }\to {x}^{\prime }{^{\mu }}=x^{\mu }-\xi ^{\mu }  \label{gg2}
\end{equation}
[this transformation changes $h$ (and hence $g)$, but of course the metric $
g_{0}$ of the background spacetime is unaffected]. The particle's worldline
in the perturbed spacetime now takes a new coordinate value,
$x'^{\mu}(\lambda )=x^{\mu }(\lambda )-\xi ^{\mu }$. Projecting now the
worldline on $g_{0}$, one obtains a new orbit ${x}^{\prime }{^{\mu }}(\tau
^{\prime })$, where $\tau ^{\prime }$ is the proper time (in $g_{0}$) of the
new orbit ${x}^{\prime }{^{\mu }}(\lambda )$. It should be emphasized that
the two projected worldlines, $x^{\mu }(\tau )$ and ${x}^{\prime }{^{\mu }}
(\tau ^{\prime })$, represent two {\em physically-distinct} trajectories in $
g_{0}$.\footnote{
Recall, however, that in the perturbed spacetime $g$ the two worldlines $
x^{\mu }(\lambda )$ and ${x}^{\prime }{^{\mu }}(\lambda )$ are physically
equivalent---they represent the same physical trajectory in two different
gauges. This difference in the relation between $x^{\mu }$ and ${x}^{\prime }
{^{\mu }}$ in the two spacetimes simply reflects the non-uniqueness of the
projection from $g$ to $g_{0}$ (which, in our ``same coordinate value''
formulation, is tied to the arbitrariness in choosing the gauge for $h$).}
In particular, the self force will now take a new value,
\begin{equation}
F_{{\rm {self}}}^{\prime \alpha }=m\left( \frac{d^{2}x^{\prime \alpha }}{
d\tau ^{\prime 2}}+\Gamma _{\mu \nu }^{\alpha }(x^{\prime })\frac{dx^{\prime
\mu }}{d\tau ^{\prime }}\frac{dx^{\prime \nu }}{d\tau ^{\prime }}\right) ,
\label{gg3}
\end{equation}
where $\Gamma _{\mu \nu }^{\alpha }(x^{\prime })$ denotes the value of the
connection in the new particle's location $x^{\prime \alpha }$.

We wish to calculate the quantity $\delta F_{{\rm {self}}}^{\alpha }$, which
is the change in $F_{{\rm {self}}}^{\alpha }$ induced by the gauge
transformation, to order $m^{2}$ [recalling that $F_{{\rm {self}}}^{\alpha }$
itself is of order $m^{2}$, and $\xi ^{\mu }$ is $O(m)$]. To this end, we
first transform the differentiation variable in Eq.\ (\ref{gg3}) from
$\tau^{\prime }$ to $\tau $:
\begin{equation}
\left( \frac{d^{2}x^{\prime \alpha }}{d\tau ^{\prime 2}}+\Gamma _{\mu \nu
}^{\alpha }(x^{\prime })\frac{dx^{\prime \mu }}{d\tau ^{\prime }}\frac{
dx^{\prime \nu }}{d\tau ^{\prime }}\right) =\left( \frac{d\tau }{d\tau
^{\prime }}\right) ^{2}\left[ \frac{d^{2}x^{\prime \alpha }}{d\tau ^{2}}
+\Gamma _{\mu \nu }^{\alpha }(x^{\prime })\frac{dx^{\prime \mu }}{d\tau }
\frac{dx^{\prime \nu }}{d\tau }\right] +\frac{d^{2}\tau }{d\tau ^{\prime 2}}
u^{\prime \alpha },  \label{gg4}
\end{equation}
where $u^{\prime \alpha }\equiv dx^{\prime \alpha }/d\tau $. Recalling that
the term in squared brackets is already $O(m)$, we may omit the factor $
(d\tau /\tau ^{\prime })^{2}=1+O(m)$, so at the required order we have
\[
F_{{\rm {self}}}^{\prime \alpha }=m\left( \frac{d^{2}x^{\prime \alpha }}{
d\tau ^{2}}+\Gamma _{\mu \nu }^{\alpha}(x^{\prime })\frac{dx^{\prime \mu }}{
d\tau }\frac{dx^{\prime \nu }}{d\tau }\right) +\beta u^{\prime \alpha },
\]
where $\beta \equiv m\frac{d^{2}\tau }{d\tau ^{\prime 2}}$. Now, the force $
F_{{\rm {self}}}^{\prime \alpha }$ must be normal to the worldline (i.e., $
F^{\prime \alpha}_{\rm self}u_{\alpha }^{\prime }=0$) by its definition in
Eq.\ (\ref{gg3}). We can therefore calculate it by projecting our last
result on the direction normal to the worldline. Noting that the term
$\beta u^{\prime \alpha }$ contributes nothing to this projection, we
obtain
\[
F_{{\rm {self}}}^{\prime \alpha }=m(\delta _{\lambda }^{\alpha }+u^{\prime
\alpha }u_{\lambda }^{\prime })\left( \frac{d^{2}x^{\prime \lambda }}{d\tau
^{2}}+\Gamma _{\mu \nu }^{\lambda}(x^{\prime})\frac{dx^{\prime \mu }}{d\tau
}\frac{dx^{\prime \nu }}{d\tau }\right) .
\]
Rewriting $F_{{\rm {self}}}^{\alpha }$ in the same form but with all primes
omitted, and subtracting it from $F_{{\rm {self}}}^{\prime \alpha }$
[evaluated at $x'(x)$], we find at order $m^{2}$
\[
\delta F_{{\rm {self}}}^{\alpha }=m(\delta _{\lambda }^{\alpha }+u^{\alpha
}u_{\lambda })\left( q^{\prime \lambda }-q^{\lambda }\right) ,
\]
where 
\[
q^{\prime \lambda }\equiv \frac{d^{2}x^{\prime \lambda }}{d\tau ^{2}}+\Gamma
_{\mu \nu }^{\lambda}(x^{\prime })\frac{dx^{\prime \mu }}{d\tau }\frac{
dx^{\prime \nu }}{d\tau }
\]
and $q^{\lambda }$ is the same but with all primes omitted. [The term
proportional to $u^{\prime \alpha }u_{\lambda }^{\prime }-u^{\alpha
}u_{\lambda }$ does not contribute at the relevant order, because it is
itself proportional to $\xi ^{\mu }$, and $q^{\lambda }$ and
$q^{\prime \lambda }$ are both $O(m)$.]
All we now need is to calculate $q^{\prime \lambda
}-q^{\lambda }$ to leading order in $\xi ^{\mu }$ [expanding $\Gamma _{\mu
\nu }^{\lambda}(x^{\prime })$ about $x^{\mu }$ to leading order in $\xi
^{\mu }$]. This is a standard calculation (it is often done when
constructing the Jacobi equation for geodesic deviation), and one finds
\[
q^{\prime \lambda }-q^{\lambda }=-\left( \ddot{\xi}^{\lambda }+{R^{\lambda }}
_{\mu \alpha \nu }u^{\mu }\xi ^{\alpha }u^{\nu }\right) ,
\]
where an overdot denotes a covariant differentiation with respect to
$\tau$ and ${R^{\lambda }}_{\mu \alpha \nu }$ is the Riemann tensor associated
with the background metric.\footnote{We use here the convention of
Ref.\ \cite{MTW} for the Riemann tensor. Notice the different convention
used by Mino {\it et al.} in \cite{MST}.}
Now, the term $u^{\alpha }u_{\lambda }$ in the above projection operator
yields vanishing contribution when applied to the term including the Riemann
tensor, due to the antisymmetry of the latter. Therefore, the final result is
\begin{equation}
\delta F_{{\rm {self}}}^{\alpha }=-m\left[ \left( g^{\alpha \lambda
}+u^{\alpha }u^{\lambda }\right) \ddot{\xi}_{\lambda }+{R^{\alpha }}_{\mu
\lambda \nu }u^{\mu }\xi ^{\lambda }u^{\nu }\right] .  \label{gg10}
\end{equation}
(Since the calculation is carried out here at order $m^{2}$ only, in the
last expression we may replace $g$ by $g_{0}$.)

The important message that arises from our discussion so far, is that the
gravitational self force is a gauge-dependent notion. Specifying $F_{\rm
self}^{\alpha}(\tau )$ by itself tells us almost nothing about the
physical self force. In order for the information on the self force to have
physical meaning, one must accompany it by the information on the gauge in
which $F_{{\rm {self}}}^{\alpha }$ was derived. Putting it in other words:
{\em The meaningful description of the gravitational self force must include
both }$F_{{\rm {self}}}^{\alpha }${\em \ and the metric perturbation} $
h_{\alpha \beta }${\em .} (Obviously, $h_{\alpha \beta }$ contains the full
information about the gauge.) This is closely related to a more general
feature of general-relativistic kinematics (in the non-perturbative
framework): Specifying the coordinate value of a worldline $x^{\mu }(\tau )$
tells one almost nothing about the physical nature of this trajectory,
unless one is also given the metric $g_{\alpha \beta }$ associated with the
coordinates $x^{\mu }$.

A remark should be made here concerning the regularity of the gravitational
self force in various gauges. The construction by MSTQW yields a regular,
well-defined, self force in the harmonic gauge. Therefore, in a given gauge
G, the self force will be well defined if and only if $\delta F_{{\rm {self}}
}^{\alpha}$ is well defined. Obviously, if the gauge transformation from H
to G is defined through a perfectly regular vector field $\xi^{\lambda}$,
the force in the G-gauge will be well defined. In most commonly used gauges,
however, the vector field $\xi^{\lambda}$ associated with the transformation
from the H-gauge to the G-gauge may inherit some of the irregularity that $
h^H$ itself possesses at the particle's location (to an extent that may
depend on the gauge G and on the physical situation). In Sec.\ \ref{secV}
this situation will be demonstrated for the RW gauge and for the radiation
gauge.

A priori it is not completely obvious what degree of regularity must be
imposed on $\xi ^{\lambda }$ in order for the self force to be regarded
``regular''. Equation (\ref{gg10}) suggests a natural criterion for
regularity: One should demand that $\xi ^{\lambda }$ will be well defined
(i.e. continuous) on the particle's worldline, and, furthermore, that along
the worldline $\xi ^{\lambda }$ will be a $C^{2}$ function of $\tau $. Note,
however, that there is some arbitrariness in choosing the regularity
criterion. For example, one might impose a stronger regularity criterion,
which requires $\xi ^{\lambda }$ to be a $C^{2}$ function of $x^{\mu }$
(such that the change in the connection due to the gauge transformation will
be well defined); but we do not see much justification for such a strong
demand. On the other hand, one may ease the above regularity criterion by
extending the standard MSTQW regularization procedure and adding to it the
element of averaging the self force (at a given moment) over all
spatial directions. With this extended procedure of regularization, one may
relax the demand for continuity of $\xi ^{\lambda }$ at the worldline,
replacing it by the weaker requirement that at the particle's location $\xi
^{\lambda }$ will have a continuous limit along each spatial geodesic
intersecting the worldline, and that this directional limit will be
integrable over the solid angle. We further discuss this possibility at the
end of the paper.

For concreteness, throughout the rest of this paper we shall adopt the
criterion which naturally follows from Eq.\ (\ref{gg10})---namely, that $\xi
^{\lambda }$ be continuous on the particle's worldline. The second half of
this criterion---the smooth dependence on $\tau $---will automatically
follow, provided that the background metric (and hence also the particle's
geodesic) is sufficiently smooth, which we assume here.\footnote{
We exclude here the situation in which the gauge condition defining the
G-gauge explicitly depends on $x^{\mu }$ or $\tau $, and this explicit
dependence artificially introduces non-smoothness to $\xi ^{\lambda }(\tau )$
. In such spurious situations we must explicitly demand that $\xi ^{\lambda
}(\tau )$ be $C^{2}$.}


\section{General gravitational forces and their gauge transformation}

\label{secIII}

The above result (\ref{gg10}) provides the full prescription for
gauge-transforming the gravitational self force. It will be instructive,
however, to address this issue of gauge transformation from yet another
point of view, by introducing the notion of a (linearized) gravitational
force and studying how this force transforms in a general gauge
transformation.

Consider again a spacetime described by a metric $g=g_{0}+h$, where $g_{0}$
is a given background metric and $h$ denotes a linearized metric
perturbation. We do {\em not} assume in this section that $h$ is a
perturbation produced by a point particle; rather, $h$ is assumed to be a
prescribed weak gravitational perturbation (it may represent, for example,
an incident gravitational wave). Suppose that a test particle with a mass $m$
is moving freely in the perturbed spacetime. Obviously, this particle will
move along a geodesic of $g$ (we neglect the self force throughout this
section\footnote{
Throughout this section we carry out the calculation to first order in the
prescribed metric perturbation $h$, {\em and to leading order in $m$} [e.g.,
order $m^{0}$ in Eq.\ (\ref{gg11}) below], so the self force is not
included.}). Namely, we shall have, in a given coordinate system $x^{\alpha }$,
\begin{equation}
\frac{d^{2}x^{\alpha }}{d\tau ^{\prime }{}^{2}}+\Gamma _{\mu \nu }^{^{\prime
}\alpha }\frac{dx^{\mu }}{d\tau ^{\prime }}\frac{dx^{\nu }}{d\tau ^{\prime }}
=0,  \label{gg11}
\end{equation}
where $x^{\alpha }(\tau ^{\prime })$ denotes the particle's trajectory in
the perturbed spacetime, $\tau ^{\prime }$ is an affine parameter (with
respect to $g$) along that trajectory, and $\Gamma _{\mu \nu }^{^{\prime
}\alpha }$ are the connection coefficients associated with the metric $g$.
However, we now wish to take the point of view according to which the
particle traces a trajectory on the background metric $g_{0}$. This
trajectory will deviate from a geodesic of the background metric $g_{0}$,
and we shall interpret this deviation as representing an external
``gravitational force'' $F_{{\rm {grav}}}^{\alpha }$, exerted on the
particle by the perturbation $h_{\alpha \beta }$. This (fictitious)
gravitational force is naturally defined as
\begin{equation}
F_{{\rm {grav}}}^{\alpha }\equiv m\ddot{x}^{\alpha }=m\left( \frac{
d^{2}x^{\alpha }}{d\tau ^{2}}+\Gamma _{\mu \nu }^{\alpha }\frac{dx^{\mu }}{
d\tau }\frac{dx^{\nu }}{d\tau }\right) ,  \label{gg12}
\end{equation}
where $\tau $ is an affine parameter in the background metric $g_{0}$, an
overdot denotes covariant differentiation (in $g_{0}$) with respect to $\tau
$, and $\Gamma _{\alpha \beta }^{\mu }$ are the connection coefficients
associated with the metric $g_{0}$. We wish to calculate $F_{{\rm {grav}}
}^{\alpha }$ to the first order in $h$ (and to the leading order in $m$).

A remark should be made here concerning the relation between the gravitational
self force and the fictitious external gravitational force considered here.
Obviously, the two notions are closely related, as both are defined through a
mapping of a worldline from the physical spacetime $g$ to a background
metric $g_{0}$. Both forces are proportional to $m$ and to the metric
perturbation $h$ (though in the self-force case one assumes that $h$ is the
metric perturbation produced by the particle itself). One may therefore be
tempted to regard the self force as a special case of the more general,
linearized gravitational force defined here. This is not quite the case,
however. The gravitational force considered here is, after all, a fictitious
force; that is, the particle actually follows a geodesic of the true
physical metric $g$ . This cannot be said about the orbit of a particle
moving under the influence of its own gravitational self force: Since the
self perturbation $h$ is singular at the particle's location, the statement
that the particle follows a geodesic of $g=g_{0}+h$ is physically
meaningless.\footnote{
One may take the point of view that the orbit of a particle under its
gravitational self force is a geodesic in a spacetime with a metric $
g_{0}+h_{{\rm {tail}}}$, where $h_{{\rm {tail}}}$ denotes the tail part of
the metric perturbation. This is, however, a fictitious geodesic, because
the actual metric is $g_{0}+h$, not $g_{0}+h_{{\rm {tail}}}$. (Recall also
that in general $h_{{\rm {tail}}}$ fails to be a vacuum solution of the
linearized Einstein equations.)} For this reason, we must view the
gravitational self force as a genuine, non-fictitious, force (though a
delicate one, as expressed by its being gauge dependent).

Proceeding with the calculation of $F_{{\rm {grav}}}^{\alpha }$, we first
transform the differentiation variable in Eq.\ (\ref{gg11}) from $\tau
^{\prime }$ to $\tau $ [mathematically this operation is the same one
applied in the previous section, Eq.\ (\ref{gg4}), though here it has a
somewhat different meaning]. We find
\begin{equation}
\frac{d^{2}x^{\alpha }}{d\tau {}^{2}}+\Gamma _{\mu \nu }^{^{\prime }\alpha }
\frac{dx^{\mu }}{d\tau }\frac{dx^{\nu }}{d\tau }+\left( \frac{d\tau ^{\prime
}}{d\tau }\right) ^{2}\frac{d^{2}\tau }{d\tau ^{\prime 2}}\frac{dx^{\alpha }
}{d\tau }=0.  \label{gg13}
\end{equation}
Denoting $\Delta \Gamma _{\mu \nu }^{\alpha }\equiv \Gamma _{\mu \nu
}^{^{\prime }\alpha }-\Gamma _{\mu \nu }^{\alpha }$ and $u^{\alpha }\equiv
dx^{\alpha }/d\tau $, and substituting Eq.\ (\ref{gg13}) in Eq.\
(\ref{gg12}) (keeping only terms linear in $h$), we obtain
\begin{equation}
F_{{\rm {grav}}}^{\alpha }=-m\Delta \Gamma _{\mu \nu }^{\alpha }u^{\mu
}u^{\nu }-\beta u^{\alpha }.  \label{gg14}
\end{equation}
We now get rid of the term $\beta u^{\alpha }$ by projecting $F_{{\rm {grav}}
}^{\alpha }$ on the subspace normal to $u^{\alpha }$, in the same way we
treated $\delta F_{{\rm {self}}}^{\alpha }$ above (recalling, again, that by
definition $F_{{\rm {grav}}}^{\alpha }$ is normal to $u^{\alpha }$). This
yields 
\begin{equation}
F_{{\rm {grav}}}^{\alpha }=-m(\delta _{\lambda }^{\alpha }+u^{\alpha
}u_{\lambda })\Delta \Gamma _{\mu \nu }^{\lambda }u^{\mu }u^{\nu }.
\label{gg15}
\end{equation}
Expressing $\Delta \Gamma $ in terms of $h$, we finally find
\begin{equation}
F_{{\rm {grav}}}^{\alpha }=-\frac{1}{2}m(g^{\alpha \lambda }+u^{\alpha
}u^{\lambda })\left( h_{\lambda \mu ;\nu }+h_{\lambda \nu ;\mu }-h_{\mu \nu
;\lambda }\right) u^{\mu }u^{\nu }.  \label{gg16}
\end{equation}
This expression (like the similar expressions below) is valid to linear
order in the perturbation $h$, and on its right-hand side we may replace $
g^{\alpha \lambda }$ by $g_{0}^{\alpha \lambda }$. It may also be useful to
express Eq.\ (\ref{gg16}) in terms of the trace-reversed metric perturbation
$\bar{h}_{\alpha \beta }\equiv h_{\alpha \beta }-\frac{1}{2}g_{\alpha \beta
}h$ (where $h\equiv g^{\alpha \beta }h_{\alpha \beta }$). One easily obtains
\begin{equation}
F_{{\rm {grav}}}^{\alpha }=mk^{\alpha \beta \gamma \delta }\bar{h}_{\beta
\gamma ;\delta }\,,  \label{gg17}
\end{equation}
where $k^{\alpha \beta \gamma \delta }$ is a tensor given by
\begin{equation}
k^{\alpha \beta \gamma \delta }=\frac{1}{2}g^{\alpha \delta }u^{\beta
}u^{\gamma }-g^{\alpha \beta }u^{\gamma }u^{\delta }-\frac{1}{2}u^{\alpha
}u^{\beta }u^{\gamma }u^{\delta }+\frac{1}{4}u^{\alpha }g^{\beta \gamma
}u^{\delta }+\frac{1}{4}g^{\alpha \delta }g^{\beta \gamma }\,.  \label{gg18}
\end{equation}

Next we investigate how this gravitational force is modified by a general
gauge transformation (\ref{gg2}). The metric perturbation $h$ transforms
according to
\[
h_{\alpha \beta }\to h_{\alpha \beta }^{\prime }=h_{\alpha \beta }+\delta
h_{\alpha \beta }\,,
\]
where 
\begin{equation}
\delta h_{\alpha \beta }=\xi _{\alpha ;\beta }+\xi _{\beta ;\alpha }\,.
\label{gg19}
\end{equation}
From Eq.\ (\ref{gg16}), the change in $h$ will induce a corresponding change
in the gravitational force $F_{{\rm {grav}}}^{\alpha }$, given by
\begin{equation}
\delta F_{{\rm {grav}}}^{\alpha }=-\frac{1}{2}m(g^{\alpha \lambda
}+u^{\alpha }u^{\lambda })\left( \delta h_{\lambda \mu ;\nu }+\delta
h_{\lambda \nu ;\mu }-\delta h_{\mu \nu ;\lambda }\right) u^{\mu }u^{\nu }.
\label{gg20}
\end{equation}

Do the self force $F_{{\rm {self}}}^{\alpha }$ and the linearized
gravitational force $F_{{\rm {grav}}}^{\alpha }$ transform in the same
manner? Substituting Eq.\ (\ref{gg19}) for $\delta h_{\alpha \beta }$ in
Eq.\ (\ref{gg20}) and using the anti-commutation relation $\xi _{\mu
;\lambda \nu }-\xi _{\mu ;\nu \lambda }=\xi _{\rho }R^{\rho }{}_{\mu \lambda
\nu }$, one obtains
\begin{eqnarray}
\delta F_{{\rm {grav}}}^{\alpha } &=&-m(g^{\alpha \lambda }+u^{\alpha
}u^{\lambda })\left( \xi _{\lambda ;\mu \nu }+\xi _{\rho }{R^{\rho }}_{\mu
\lambda \nu }\right) u^{\mu }u^{\nu }  \nonumber \\
&=&-m\left[ \left( g^{\alpha \lambda }+u^{\alpha }u^{\lambda }\right) \ddot{
\xi}_{\lambda }+{R^{\alpha }}_{\mu \lambda \nu }u^{\mu }\xi ^{\lambda
}u^{\nu }\right] .  \label{gg30a}
\end{eqnarray}
Comparing this expression to Eq.\ (\ref{gg10}), we find that the two forces
admit the same transformation law:
\begin{equation}
\delta F_{{\rm {grav}}}^{\alpha }=\delta F_{{\rm {self}}}^{\alpha }.
\label{gg300}
\end{equation}
This result is not surprising, because the two types of forces share a
common kinematic feature: They are both constructed through a projection of
a worldline from a physical metric $g$ to a background metric $g_{0}$, and
therefore they transform in the same manner.


\section{Regularizing the gravitational self force in various gauges}

\label{secIV}

The method developed by MSTQW for regularizing the gravitational self force
is formulated within the framework of the harmonic gauge. This means that in
Eq.\ (\ref{Int-1}) above, the two quantities on the right-hand side, $
F^{\alpha {\rm {(bare)}}}$ and $F^{\alpha {\rm {(inst)}}}$ are to be
evaluated in the harmonic gauge---and the outcome is the self force {\em in
the harmonic gauge}. We therefore rewrite this equation explicitly as
\begin{equation}
F_{{\rm {self}}}^{{\rm {(H})}}=F_{\rm bare}^{\rm (H)}-F_{\rm inst}^{\rm (H)},
\label{gg31}
\end{equation}
where the parenthetical index ``H'' denotes the harmonic gauge (for
brevity we omit the tensorial index $\alpha $ here and in the equations
below).

Assume now that a gauge transformation is made, from the harmonic gauge to a
new gauge which we denote schematically by ``G''. According to the
discussion in Sec. \ref{secII}, the self force in the new gauge will be
given by
\[
F_{{\rm {self}}}^{{\rm {(G)}}}{}=F_{{\rm {self}}}^{{\rm {(H)}}}{}+\delta F_{{\rm
{self}}}^{{\rm {(H\rightarrow G)}}}=\left[ \delta F_{{\rm {self}}}^{{\rm {\
(H\rightarrow G)}}}+F_{\rm bare}^{\rm (H)}\right] -F_{\rm inst}^{\rm (H)},
\]
where $\delta F_{{\rm {self}}}^{{\rm {(H\rightarrow G)}}}$ is the expression
given in Eq.\ (\ref{gg10}), with $\xi ^{\lambda }$ being the displacement
vector that transforms from the harmonic gauge to the new gauge G. To
evaluate the term in squared brackets, we first recall that the ``bare
force'' is related to the trace-reversed metric perturbation through
\begin{equation}
F^{\alpha}_{\rm bare}=mk^{\alpha \beta \gamma \delta }\bar{h}_{\beta
\gamma ;\delta }  \label{gg34}
\end{equation}
[see the second equality in Eq.\ (28) of Ref.\ \cite{paperI}], which is
expressed in terms of the metric perturbation itself as
\begin{equation}
F^{\alpha}_{\rm bare}=-\frac{1}{2}m(g^{\alpha \lambda }+u^{\alpha
}u^{\lambda })\left( h_{\lambda \mu ;\nu }+h_{\lambda \nu ;\mu }-h_{\mu \nu
;\lambda }\right) u^{\mu }u^{\nu }.  \label{gg35}
\end{equation}
Noting further that Eqs.\ (\ref{gg20}) and (\ref{gg300}) imply
\begin{equation}
\delta F_{{\rm {self}}}^{\alpha }=-\frac{1}{2}m(g^{\alpha \lambda
}+u^{\alpha }u^{\lambda })\left( \delta h_{\lambda \mu ;\nu }+\delta
h_{\lambda \nu ;\mu }-\delta h_{\mu \nu ;\lambda }\right) u^{\mu }u^{\nu },
\label{gg39}
\end{equation}
we then obtain (using $h^{{\rm {(H)}}}+\delta h^{{\rm {(H\rightarrow G)}}}
=h^{{\rm {(G)}}}$)
\begin{equation}
F_{\rm bare}^{(H)}+\delta F_{{\rm {self}}}^{{\rm {(H\rightarrow G)}}}=
F_{\rm bare}^{(G)}.  \label{gg40}
\end{equation}
This result has a simple interpretation in terms of the notion of
``gravitational force'' discussed in the previous section: (i) As was
established there, the self force and the gravitational force transform
exactly in the same manner, and (ii) the ``bare force'' is nothing but the
gravitational force associated with the full metric perturbation $h$
(produced by the particle). It then follows that the self force and the bare
force transform in the same manner.\footnote{
It should be emphasized that the physical notion of ``gravitational force''
introduced in the previous section is not necessary for the derivation of
Eq.\ (\ref{gg40}). Thus, starting from Eq.\ (\ref{gg10}), one can derive
Eq.\ (\ref{gg39}) directly as a mathematical identity [following the same
mathematical steps used above for constructing Eq.\ (\ref{gg30a}) from
Eq.\ (\ref{gg20})], without any reference to the notion of ``gravitational
forces''.}

We conclude that in an arbitrary gauge G the regularized gravitational self
force is simply given by
\begin{equation}  \label{gg45}
F_{{\rm {self}}}^{{\rm {(G)}}}=F_{\rm bare}^{(G)}-F_{\rm inst}^{(H)}.
\end{equation}
Namely, in an arbitrary gauge G, the singular piece to be subtracted from
the bare force is always the instantaneous piece expressed {\em in the
harmonic gauge}, and not in the gauge G, as one might naively expect.

Our last result is of special importance: The analysis by MSTQW tells us how
to calculate the physical self force associated with the metric perturbation
in the harmonic gauge. In particular, it tells us how to construct the
``correct'' instantaneous part of the bare force in this gauge. Our above
discussion implies that even when calculating the self force in a different
gauge, the ``correct'' instantaneous part must still be calculated in the
harmonic gauge. [The explicit construction of the instantaneous part from
the harmonic gauge Green's function is described in Eq.\ (29) of Ref.\ \cite
{paperI}]. This harmonic-gauge-related instantaneous part is the one which
captures the ``correct'' divergent piece to be removed from the bare force
in whatever gauge. Intuitively, this special significance of the harmonic
gauge may be attributed to its inherently isotropic nature: The ``correct''
divergent piece that should be removed from the bare force must be spatially
isotropic (see, e.g., the analysis by Quinn and Wald \cite{QW}), and it is
the harmonic gauge which admits this isotropic structure; other gauge
conditions may introduce an artificial distortion to the singular piece.

\subsection*{Mode-sum regularization in various gauges}

In Ref.\ \cite{paperI} we introduced a practical calculation scheme for the
gravitational self force, based on the regularization procedure by MSTQW,
which employs a multipole mode decomposition. This method of ``mode sum
regularization'' has been formulated in Ref.\ \cite{paperI} only within the
harmonic gauge. Let us now examine how the above discussion, concerning the
construction of the regularized self force in various gauges, applies in the
framework of the mode-sum scheme.

Within the mode sum scheme, the harmonic-gauge regularized gravitational
self force is given by \cite{paperI}
\begin{equation}
F^{\alpha{\rm (H)}}_{\rm self}=\sum_{l=0}^{\infty }
\left(F^{\alpha l{\rm (H)}}_{\rm bare}-A^{\alpha }L-B^{\alpha }-C^{\alpha }/L\right)
-D^{\alpha },  \label{eqVI160}
\end{equation}
where the summation is over multipole modes $l$, and $L\equiv l+1/2$. In
this expression, $F^{\alpha l{\rm (H)}}_{\rm bare}$ is the
contribution to the self force from $h_{\beta\gamma}^{l{\rm (H)}}$, the
$l$-mode of the metric perturbation in the harmonic gauge. This contribution
is given by
\begin{equation}
F^{\alpha l{\rm (H)}}_{\rm bare}=mk^{\alpha\beta \gamma\delta}
\bar{h}_{\beta\gamma;\delta}^{l{\rm (H)}},  \label{eqVI20}
\end{equation}
where $\bar{h}_{\beta\gamma}^{l{\rm (H)}}$ is the trace-reversed $
h_{\beta\gamma}^{l{\rm (H)}}$ and $k^{\alpha\beta\gamma\delta}$ is
the tensor given in Eq.\ (\ref{gg18}). The vectorial quantities $A^{\alpha
} $, $B^{\alpha }$, $C^{\alpha }$, and $D^{\alpha }$ appearing in Eq.\ (\ref
{eqVI160}) are $l$-independent. These quantities, which we call
``regularization parameters'', are constructed from the $l$-modes of the
instantaneous part $F_{\rm inst}^{(H)}$, in a manner described in Ref.\
\cite{paperI}.

The prescription provided by Eq.\ (\ref{eqVI160}) yields the ``harmonic
gauge'' self force. It is now possible, however, to re-formulate this
prescription in any other gauge ``G'', using
\begin{equation}
F^{\alpha{\rm (G)}}_{\rm self}=F^{\alpha{\rm (H)}}_{\rm self}
+\delta F^{\alpha{\rm (H\rightarrow G)}}_{\rm self}.
\label{161}
\end{equation}
Rewriting Eq.\ (\ref{gg39}) as $\delta F_{{\rm {self}}}^{\alpha }=mk^{\alpha
\beta \gamma \delta }\delta \bar{h}_{\beta \gamma ;\delta }$ (where $\delta
\bar{h}$ denotes the trace-reversed $\delta h$), and decomposing $\delta
\bar{h}$ into $l$-modes, we obtain
\[
\delta F^{\alpha{\rm (H\rightarrow G)}}_{\rm self}
=\sum_{l=0}^{\infty }mk^{\alpha\beta\gamma\delta}
\delta \bar{h}_{\beta\gamma;\delta }^{l{\rm (H\rightarrow G)}}.
\]
Substituting this and Eq.\ (\ref{eqVI160}) into Eq.\ (\ref{161}), we find
\begin{equation}
F^{\alpha{\rm (G)}}_{\rm self}=\sum_{l=0}^{\infty }\left[ \left(
F^{\alpha l{\rm (H)}}_{\rm bare}+mk^{\alpha\beta \gamma \delta}
\delta \bar{h}_{\beta \gamma ;\delta }^{l{\rm (H\rightarrow G)}}
\right) -A^{\alpha }L-B^{\alpha }-C^{\alpha }/L\right] -D^{\alpha }.
\end{equation}
Using now Eq.\ (\ref{eqVI20}), we can re-express the term in parentheses
as
\[
mk^{\alpha\beta\gamma\delta}\bar{h}_{\beta\gamma;\delta}^{l{\rm (H)}}
+mk^{\alpha\beta\gamma\delta}\delta\bar{h}_{\beta
\gamma;\delta}^{l{\rm (H\rightarrow G)}}=mk^{\alpha\beta
\gamma \delta}\bar{h}_{\beta\gamma;\delta}^{l{\rm (G)}}\equiv
F^{\alpha l{\rm (G)}}_{\rm bare},
\]
where $F^{\alpha l{\rm (G)}}_{\rm bare}$ denotes $l$-mode
contribution to the ``G-gauge bare force'', namely, the contribution to the
bare force from the mode $l$ of the (bare) metric perturbation in the
G-gauge, through Eq.\ (\ref{gg34}). We thus obtain the simple expression
for the self force in an arbitrary gauge ``G'',
\begin{equation}
F^{\alpha{\rm {(G)}}}_{\rm self}=\sum_{l=0}^{\infty }\left(F^{\alpha
l{\rm (G)}}_{\rm bare}-A^{\alpha }L-B^{\alpha }-C^{\alpha}/L\right)
-D^{\alpha }.  \label{eqVI170}
\end{equation}

We conclude that the regularization parameters $A^{\alpha }$, $B^{\alpha }$,
$C^{\alpha }$, and $D^{\alpha }$ are independent of the gauge. This result
has a simple intuitive explanation: These parameters are determined by
the mode decomposition of the {\em instantaneous} piece of the metric
perturbation,
which---based on our above discussion---is always to be expressed in the
harmonic gauge, regardless of the gauge chosen for calculating the self
force. Thus, the regularization parameters $A^{\alpha }$ , $B^{\alpha }$, $
C^{\alpha }$, and $D^{\alpha }$ are, in effect, {\em gauge-independent}.

It should be commented that the above discussion is valid as long as $
\delta F^{\alpha{\rm (H\rightarrow G)}}_{\rm self}$ (and hence
the self force in the gauge ``G'') admits a well defined finite value. As we
demonstrate in the next section, in certain gauges the self force turns out
to be irregular or ill-defined. In such cases, the irregularity may enter
Eq.\ (\ref{eqVI170}) through the bare modes $F^{\alpha l{\rm (G)}}_{\rm bare}$
and render the sum over $l$ non-convergent.


\section{Examples}

\label{secV}

In this section we study the transformation of the self force from the
harmonic gauge to other, commonly used gauges, in a few simple cases. In
principle, this transformation is done by first solving Eq.\ (\ref{gg19})
for the gauge displacement vector $\xi ^{\mu }$, and then constructing the
force difference $\delta F_{{\rm self}}^{\alpha }$ by using Eq.\ (\ref{gg10}).
We shall primarily be concerned here about the regularity of the self
force in the new gauge. As discussed in Sec.\ \ref{secII}, we shall regard
the G-gauge self force as regular if the vector field $\xi ^{\mu }$ is
continuous at the particle's location. If it is indeed continuous, then the
self force in the new gauge is given in Eq.\ (\ref{gg10}) (the demand for a $
C^{2}$ dependence on $\tau $ is automatically satisfied, as discussed in
Sec.\ \ref{secII}). We begin by considering the transformation to the
Regge-Wheeler (RW) gauge, for radial trajectories in the Schwarzschild
spacetime. Then we examine the transformation to the RW gauge for a uniform
circular orbit. Finally, we examine the transformation to the (outgoing)
radiation gauge, in a simple flat-space example.

\subsection{Regge--Wheeler gauge: radial trajectories}

We consider a particle of mass $m$ moving along a strictly radial free-fall
orbit on the background of a Schwarzschild black hole with mass $M\gg m$.
(Of course, the motion of the particle will remain radial even under the
effect of self-force, by virtue of the symmetry of the problem.) In what
follows we use Schwarzschild coordinates $t,r,\theta ,\varphi $ and
assume, without loss of generality, that
the radial trajectory lies along the polar axis, i.e., at $\theta =0$.

Let $h_{\alpha \beta }^{{\rm (H)}}$ and $h_{\alpha \beta }^{{\rm (RW)}}$ denote
the metric perturbation produced by the above particle in the harmonic and
RW gauges, respectively. The displacement vector field $\xi ^{\mu }$ which
transforms $h_{\alpha \beta }^{{\rm (H)}}$ to $h_{\alpha \beta }^{{\rm (RW)}}$
satisfies the gauge transformation equation
\begin{equation}
h_{\alpha \beta }^{{\rm (RW)}}=h_{\alpha \beta }^{{\rm (H)}}
+\xi _{\alpha ;\beta}+\xi _{\beta ;\alpha }.  \label{eqVI180}
\end{equation}
The symmetry of the physical setup motivates one to consider only
axially-symmetric even-parity metric perturbation modes. Accordingly, we
shall look for solutions to Eqs.\ (\ref{eqVI180}) which are
$\varphi$-independent and also have $\xi _{\varphi }=0$.

For even-parity perturbation modes, the RW gauge conditions take the simple
algebraic form \cite{RW}
\begin{equation}
h_{t\theta }^{{\rm (RW)}}=h_{r\theta }^{{\rm (RW)}}=h_{{\rm ang}}^{{\rm (RW)}}=0,
\label{eqVI190}
\end{equation}
where $h_{{\rm ang}}\equiv \left( h_{\theta \theta }-\sin ^{-2}\theta
\,h_{\varphi \varphi }\right) /2$. Imposing these conditions, the gauge
transformation equation (\ref{eqVI180}) yields three coupled differential
equations for the three components $\xi _{t}$, $\xi _{r}$, and $\xi _{\theta
}$: 
\begin{mathletters}
\label{eqVI200}
\begin{equation}
\xi _{t,\theta }+\xi _{\theta ,t}=-h_{t\theta }^{{\rm (H)}}\,,
\label{eqVI200a}
\end{equation}
\begin{equation}
\xi _{r,\theta }+\xi _{\theta ,r}-(2/r)\xi _{\theta }=-h_{r\theta }^{{\rm (H)}
}\,,  \label{eqVI200b}
\end{equation}
\begin{equation}
\sin \theta (\sin ^{-1}\theta \,\xi _{\theta })_{,\theta }=-h_{{\rm ang}}^{
{\rm (H)}}\,.  \label{eqVI200d}
\end{equation}
Eq.\ (\ref{eqVI200d}) can be immediately integrated with respect to $\theta $
(with fixed $t,r$), yielding
\end{mathletters}
\begin{equation}
\xi _{\theta }=-\sin \theta \left[ \int_{0}^{\theta }\sin ^{-1}\theta
^{\prime }\,h_{{\rm ang}}^{{\rm (H)}}d\theta ^{\prime }+\psi _{1}(r,t)\right] ,
\label{eqVI205}
\end{equation}
where $\psi _{1}$ is an arbitrary function. (As we shall discuss below, $h_{
{\rm ang}}^{{\rm (H)}}$ vanishes sufficiently fast as $\theta ^{\prime
}\rightarrow 0$, such that the integral is well-defined at the lower limit.)
Then, Eqs.\ (\ref{eqVI200a}) and (\ref{eqVI200b}) are immediately
solvable, yielding
\begin{equation}
\xi _{t}=-\int_{0}^{\theta }\left( h_{t\theta }^{{\rm (H)}}+\xi _{\theta
,t}\right) d\theta ^{\prime }+\psi _{2}(r,t),\quad \quad \xi
_{r}=-\int_{0}^{\theta }\left[ h_{r\theta }^{{\rm (H)}}+\xi _{\theta
,r}-(2/r)\xi _{\theta }\right] d\theta ^{\prime }+\psi _{3}(r,t),
\label{eqVI206}
\end{equation}
where $\psi _{2}$ and $\psi _{3}$ are two other arbitrary
functions.\footnote{
The arbitrary functions $\psi _{i}$ represent a true freedom in the
construction of the RW-gauge metric perturbations.
This may be attributed to the freedom of specifying the monopole and dipole
modes of the metric perturbation---see the discussion in Ref. \cite{RW}.}

Now, in order to explore the behavior of the quantity $\delta F_{{\rm self}
}^{\alpha }$ corresponding to the gauge transformation $\text{H$\to $ RW}$,
by means of Eq.\ (\ref{gg10}), one has to characterize the behavior of
the vector field $\xi ^{\mu }$ at the location of the particle.
This requires one to
first explore the behavior of the various H-gauge metric functions appearing
in Eqs.\ (\ref{eqVI205}) and (\ref{eqVI206}) at the particle's location.
This task is most easily accomplished by considering the Hadamard form of
the metric perturbation in the neighborhood of the particle. For the
trace-reversed metric perturbation in the harmonic gauge, this form was
given by Mino {\it et al.} [see Eq.\ (2.27) of Ref.\ \cite{MST};
Alternatively, see Eq.\ (45) of Ref.\ \cite{QW}]:\footnote{
To obtain Eq.\ (\ref{eqVI210}) from Eq.\ (2.27) of Ref.\ \cite{MST}, recall
that at the location of the particle we have $\bar g^{\mu}{}_{\alpha}=
\delta^{\mu}_{\alpha}$ 
and $\kappa= 1$ 
(using the notation of \cite{MST}).}
\begin{equation}
\bar{h}_{\alpha \beta }^{\rm (H)}=4m\epsilon ^{-1}u_{\alpha
}u_{\beta }+O(\epsilon ^{0}),  \label{eqVI210}
\end{equation}
where $\epsilon $ is the spatial geodesic distance to the particle's
worldline (i.e., the proper length of the geodesic normal to the worldline
which connects the latter to the evaluation point), and the terms included
in $O(\epsilon ^{0})$ are assured to be at least $C^{1}$ functions of the
coordinates at $\epsilon =0$. The metric perturbation itself is then given
by 
\begin{equation}
h_{\alpha \beta }^{\rm (H)}=4m\epsilon ^{-1}\left( u_{\alpha
}u_{\beta }+g_{\alpha \beta }/2\right) +O(\epsilon ^{0}).  \label{eqVI220}
\end{equation}

Since the worldline is radial (namely $u_{\theta }=u_{\varphi }=0$), it now
follows that the metric perturbation components $h_{t\theta }^{{\rm H}}$, $
h_{r\theta }^{{\rm (H)}}$, and $h_{{\rm ang}}^{{\rm (H)}}$ appearing in Eqs.\
(\ref{eqVI205}) and (\ref{eqVI206}) all have vanishing contributions from the
singular $O(\epsilon ^{-1})$ term, and are therefore all regular (i.e., at
least $C^{1}$) on the worldline. Consequently, one can easily construct
solutions for $\xi ^{\mu }$, which have regular, finite values at the
particle's location: Starting from Eq.\ (\ref{eqVI205}), we first observe
(e.g., by transforming to cartesian-like coordinates at the polar axis, and
demanding axial symmetry as well as $C^{1}$ asymptotic behavior at $\theta
=0 $) that $h_{{\rm ang}}^{{\rm (H)}}$ falls off at $\theta \to 0$ faster than
$\theta $. As a consequence, the integral in Eq.\ (\ref{eqVI205}), too,
falls off faster than $\theta $. Thus, $\xi _{\theta }$ is regular at $
\theta \to 0 $, and it vanishes there like $\propto \theta $. (With the
choice $\psi _{1}=0$, $\xi _{\theta }$ would vanish even faster than $\theta
^{2}$).

Consider next the two integrals in Eq.\ (\ref{eqVI206}). From the above
discussion it immediately follows that the two derivatives $\xi _{\theta ,t}$
and $\xi _{\theta ,r}$ vanish like $\propto \theta $ (at least)---like $\xi
_{\theta }$ itself. Since $h_{t\theta }^{{\rm H}}$ and $h_{r\theta }^{{\rm H}
}$ are regular ($C^{1}$) too, we find that the two integrands in Eq.\ (\ref
{eqVI206}) are bounded at $\theta =0$. (In fact, by transforming to
cartesian-like coordinates near $\theta =0$ one can easily verify that $
h_{t\theta }^{{\rm H}}$ and $h_{r\theta }^{{\rm H}}$---and hence the two
integrands---vanish at $\theta \to 0$.) Consequently, the two integrals
vanish at $\theta \to 0$. We find that along the particle's worldline all
components of $\xi ^{\mu }$ are regular, and satisfy
\[
\xi _{\theta }=0,\quad\quad
\xi _{t}=\psi _{2}(r,t),\quad\quad
\xi _{r}=\psi _{3}(r,t),
\]
where $\psi _{2}(r,t)$ and $\psi _{3}(r,t)$ are freely-specifiable
functions. (In fact, this holds not only at the particle's worldline, but
everywhere along the polar axis.) Furthermore, choosing $\psi _{2}=\psi
_{3}=0$, we obtain a solution for $\xi ^{\mu }$ which is not only regular
but is also vanishing along the particle's worldline: $\xi ^{\mu }(\tau)=0$.

Since the above-constructed vector $\xi ^{\mu }$ is continuous at the
particle's location, we obtain---through Eq.\ (\ref{gg10})---a regular
finite value for the desired quantity $\delta F_{{\rm self}}^{\alpha }$.
Thus, {\em for strictly radial trajectories in Schwarzschild spacetime, the
gravitational self force is regular in the RW gauge}. Moreover, this
RW-gauge self force can be made equal to the harmonic-gauge self force, by
exploiting the remaining freedom in the RW gauge (manifested here by the
three arbitrary functions $\psi _{1-3}$).

\subsection{Regge--Wheeler gauge: circular orbits}

Let us now consider a particle which (in the lack of self force) moves on a
circular geodesic at $r=r_{0}\geq 6M$ around a Schwarzschild black hole.
Without loss of generality, we shall assume an equatorial orbit (i.e.,
$\theta =\pi/2$ and $u^{\theta }=0$) and will consider the self force at
a point P located on the particle's orbit at $t=\varphi =0$.
In this physical scenario, the metric perturbation contains both even and
odd parity modes. The RW gauge condition \cite{RW} then becomes a bit more
complicate than the one specified in Eq.\ (\ref{eqVI190}) for a purely even
perturbation (in general, the two algebraic conditions $h_{t\theta }^{{\rm (RW)
}}=h_{r\theta }^{{\rm (RW)}}=0$ are no longer valid, and are to be replaced by
conditions involving derivatives of the metric perturbation). However, the
two gauge conditions involving the angular components of the metric
perturbation maintain a simple algebraic form, namely
\begin{equation}
h_{\theta \varphi }^{{\rm (RW)}}=0,\quad \quad h_{{\rm ang}}^{{\rm (RW)}}=0.
\label{gg80}
\end{equation}
For our purpose, it will be sufficient to consider only these two
conditions. When imposed on the gauge transformation equation
(\ref{eqVI180}), these conditions lead to a set of two coupled equations
for $\xi _{\theta}$ and $\xi _{\varphi }$:
\begin{mathletters}
\label{gg90}
\begin{equation}
\sin \theta (\sin ^{-1}\theta \,\xi _{\theta })_{,\theta }-\sin ^{-2}\theta
\,\xi _{\varphi ,\varphi }=-h_{{\rm ang}}^{{\rm (H)}}.  \label{gg90b}
\end{equation}
\begin{equation}
\xi _{\theta ,\varphi }+\sin ^{2}\theta (\sin ^{-2}\theta \,\xi _{\varphi
})_{,\theta }=-h_{\theta \varphi }^{{\rm (H)}},  \label{gg90a}
\end{equation}
\end{mathletters}
The source terms for these equations are evaluated, again, with the help of
Eq.\ (\ref{eqVI220}): We find that $h_{\theta \varphi }^{{\rm (H)}}$ is
regular at the particle's location, but the source for Eq.\ (\ref{gg90b})
diverges there as
$-h_{{\rm ang}}^{{\rm (H)}}\cong a\, r_0^{2}\epsilon ^{-1}$,
where $a=2mr_0^{-2}u_{\varphi}^2=2m(r_0/M-3)^{-1}$ \cite{Chandra}
and, as before, $\epsilon $ denotes the spatial geodesic distance to the
particle's worldline. In what follows we analyze the behavior of $\xi
_{\theta }$ and $\xi _{\varphi }$ at the immediate neighborhood of P, to
leading order in $\epsilon $.

We first note that no derivatives with respect to $r$ and $t$ appear in Eq.\
(\ref{gg90b}) (though the source term depends on $r$ and $t$ through $
\epsilon $). Therefore, this equation can be solved for each $r,t$
separately. For our purpose---demonstrating the discontinuity of the
solution at P---it will be sufficient to consider the solution at the
two-dimensional plain $r=r_{0},t=0$, which is simpler to analyze.

To bring Eqs.\ (\ref{gg90}) to a convenient form, we introduce the local
cartesian-like coordinates $y\equiv r_{0}\sin \theta \sin \varphi $, $
z\equiv r_{0}\cos \theta $ in the neighborhood of P.
Note that $z=y=0$ at P, and that (for $r=r_{0},t=0$) at the leading
order we have $\epsilon =[(1-v^{2})^{-1}y^{2}+z^{2}]^{1/2}$.
Here $v$ denotes the particle's velocity in the Lorentz frame of a
static local observer,
$
v\equiv (-g_{\varphi \varphi }/g_{tt})^{1/2}(d\varphi /dt)
$.
One can easily obtain the explicit value of $v$: \cite{Chandra}
\[
v=(-g^{\varphi \varphi }/g^{tt})^{1/2}(u_{\varphi
}/u_{t})=(r_{0}/M-2)^{-1/2} <1 .
\]
Transforming in Eqs.\
(\ref{gg90}) from $(\theta ,\varphi)$ to $(z,y)$ we obtain two coupled
equations for $\xi _{z}$ and $\xi _{y}$, reading
\begin{mathletters}
\label{gg100}
\begin{equation}
(1-z^{2}/r_{0}^{2})\xi _{z,z}-(1-y^{2}/r_{0}^{2})\xi _{y,y}=a/\epsilon
+\cdots ,  \label{gg100a}
\end{equation}
\begin{equation}
\xi _{z,y}+\xi _{y,z}-2yz(r_{0}^{2}-z^{2})^{-1}\xi _{y,y}=0+\cdots ,
\label{gg100b}
\end{equation}
where the
dots ($\cdots $) represent corrections to the source term which are at least
$C^{1}$ at P. As we are interested only in the leading-order behavior of $
\xi ^{\mu }$ at P [where $(z/r_{0})^{2},(y/r_{0})^{2},(zy/r_{0}^{2})\ $all
vanish], we shall proceed by restricting attention to the leading-order form
of Eqs.\ (\ref{gg100}):
\end{mathletters}
\begin{mathletters}
\label{gg101}
\begin{equation}
\xi _{z,z}-\xi _{y,y}=a/\epsilon ,  \label{gg101a}
\end{equation}
\begin{equation}
\xi _{z,y}+\xi _{y,z}=0.  \label{gg101b}
\end{equation}
Equation (\ref{gg101b}) allows us to express the vector $\xi ^{\mu }$ in
terms of a scalar potential $\Phi $, as \footnote{
Defining $\vec{E}\equiv (E_{y},E_{z})\equiv (-\xi _{y},\xi _{z})$, Eq.\ (\ref
{gg101b}) reads $\nabla \times \vec{E}{}=0$, which allows one to define ${}
\vec{E}=\nabla \Phi $. (Note, however, that since there is a singularity at $
y=z=0$, $\Phi $ needs not be single-valued -- see the discussion below).}
\end{mathletters}
\begin{equation}
\xi _{z}=\Phi _{,z}\quad ,\quad \xi _{y}=-\Phi _{,y}\,.  \label{gg102}
\end{equation}
With Eq.\ (\ref{gg101a}), this potential is then found to satisfy Poisson's
equation
\begin{equation}
\Phi _{,zz}+\Phi _{,yy}=a/\epsilon .  \label{gg103}
\end{equation}

It is convenient to introduce polar coordinates in the $zy$-plain, which we
define through $z=\rho \sin \phi $, $y=\rho \cos \phi $. Transforming in
Eq.\ (\ref{gg103}) from $z,y$ to $\rho ,\phi $, and substituting $\epsilon
=[(1-v^{2})^{-1}y^{2}+z^{2}]^{1/2}$, we obtain
\begin{equation}
\rho ^{-1}\left( \rho \Phi _{,\rho }\right) _{,\rho }+\rho ^{-2}\Phi _{,\phi
\phi }=\frac{a}{\rho }(1-v^{2})^{1/2}\left( 1-v^{2}\sin ^{2}\phi \right)
^{-1/2}.  \label{gg105}
\end{equation}

Next, we wish to expand $\Phi (\rho ,\phi)$ into angular Fourier modes $
e^{in\phi }$. Before doing this, however, there is a subtlety that must be
discussed. The displacement vector $\xi ^{\mu }$ must be a single-valued
(SV) function of $\phi $. This means that both $\Phi _{,\phi }$ and $\Phi
_{,\rho }$ must be SV too. However, in principle the generating potential $
\Phi $ needs {\em not} be a SV function of $\phi $. Therefore, in the
complete mode decomposition of $\Phi $ one may also include certain
functions of $\phi $ which are not necessarily SV. However, since the
$\phi $-derivative of each such multi-valued function must be SV, this function
must be {\em linear} in $\phi $ (such that the Fourier expansion of $\Phi
_{,\phi }$ will only include SV Fourier modes). Furthermore, since the $\rho
$-derivative must be SV too, this ``linear mode'' must be independent of $
\rho $. The full decomposition thus takes the form
\begin{equation}
\Phi (\rho ,\phi)=c\phi +\sum_{n=-\infty }^{\infty }e^{in\phi }\Phi
_{n}(\rho)\,,  \label{gg106}
\end{equation}
where $c$ is an arbitrary constant. Substituting this form in Eq.\ (\ref
{gg105}) (recalling that $c\phi $ satisfies the homogeneous part of this
equation), one obtains an ordinary equation for each $n$-mode, reading
\begin{equation}
\rho ^{-1}\left( \rho \Phi _{n,\rho }\right) _{,\rho }-n^{2}\rho ^{-2}\Phi
_{n}=\frac{a}{\rho }\,f_{n}\,,  \label{gg108}
\end{equation}
where the coefficients $f_{n}$ are given by the (elliptic) integrals
\begin{equation}
f_{n}=\frac{\sqrt{1-v^{2}}}{2\pi }\int_{0}^{2\pi }\frac{e^{-in\phi }}{\sqrt{
1-v^{2}\sin ^{2}\phi }}\,d\phi .  \label{gg109}
\end{equation}
It can be easily verified that
$f_{n}$ vanishes for all odd $n$. For even $n$, however, $f_{n}$ is
generally non-vanishing. In particular, for $n=0$ the integrand in Eq.\ (\ref
{gg109}) is bounded from below by unity, hence $f_{0}>\sqrt{1-v^{2}}>0$.

The general exact solution to Eq.\ (\ref{gg108}) is easily constructed:
\begin{equation}
\Phi _{n}=\left\{
\begin{array}{ll}
b_{0}\rho +\alpha _{0}+\beta _{0}\ln \rho , & \text{for $n=0$}, \\
b_{n}\rho +\alpha _{n}\rho ^{|n|}+\beta _{n}\rho ^{-|n|}, & \text{for $n\ne
0 $},
\end{array}
\right.  \label{gg110}
\end{equation}
where $\alpha _{n}$ and $\beta _{n}$ are arbitrary constants, and
\[
b_{n}=\left\{
\begin{array}{ll}
af_{n}/(1-n^{2}) & \text{for even $n$}, \\
0 & \text{for odd $n.$}
\end{array}
\right.
\]
We may now construct the modes of $\xi ^{\mu }$ by applying Eq.\
(\ref{gg102}) to each of the single modes. We then wish to figure out
what is the
solution with the most regular behavior at the limit $\rho \to 0$, which
concerns as here. Clearly, any choice of $\beta _{n}\neq 0$ will lead to a
divergent $\Phi _{n}$ and hence to a divergent vector $\xi ^{\mu }$. [Note
that the norm of $(\xi _{y},\xi _{z})$ is the same as that of $\nabla \Phi $
, and is hence bounded below by $|\Phi _{,\rho }|$; and the contribution to
the latter from a nonvanishing $\beta _{n}$ would diverge like $\propto \rho
^{-|n|-1}$.] Similarly, a nonvanishing $c$ would yield a potential $\Phi $
whose (normalized) derivative in the tangential direction, $\rho ^{-1}\Phi
_{,\phi }$, diverges like $c\rho ^{-1}$.\footnote{
Divergent contributions from different $n$-modes cannot cancel each other,
because they have different dependence on $\phi $, as well as different
rates of divergence ($\rho ^{-|n|-1}$). Also, a divergence coming from the
linear mode cannot cancel a divergent $n=0$ mode, even though in both modes $
|\nabla \Phi |\propto \rho ^{-1}$, because the direction of $\nabla \Phi $
is tangential for the linear mode and ``radial'' for the $n=0$ mode.} The
most regular solution is thus one with $\beta _{n}=0$ for all $n$, as well
as $c=0$. This solution takes the form
\[
\Phi (\rho ,\phi)=\alpha_0+\rho H(\phi)+O(\rho ^{2})\,,
\]
where
\[
H(\phi)=\left( \alpha _{1}e^{i\phi }+\alpha _{-1}e^{-i\phi }\right)
+\sum_{n=-\infty }^{\infty }b_{n}e^{in\phi }.
\]

Returning from $\Phi $ to $\xi ^{\mu }$, we find e.g. for the cartesian-like
component $\xi _{y}$ (ignoring higher-order contributions in $\rho $):
\[
\xi _{y}=-\Phi _{,y}=-\rho _{,y}H-\rho \phi _{,y}H_{,\phi }\,.
\]
Substituting $\rho _{,y}=\cos \phi $ and $\phi _{,y}=-\rho ^{-1}\sin \phi $,
we find
\[
\xi _{y}=-H\cos \phi +H_{,\phi }\sin \phi \equiv \xi _{y}(\phi )\,\,.
\]
Clearly, in order for $\xi _{y}$ to be continuous at $\rho \to 0$ (where $
\phi $ is indefinite), it must be independent of $\phi $. However, $\xi
_{y,\phi }=(H_{,\phi \phi }+H)\sin \phi $, and
\[
H_{,\phi \phi }+H=\sum_{n=-\infty }^{\infty }(1-n^{2})b_{n}e^{in\phi
}\,=a\sum_{n=-\infty }^{\infty }f_{n}e^{in\phi }.
\]
This function of $\phi $ does not vanish (identically) unless all
coefficients $f_{n}$ vanish; however, as was shown above, $f_{0}\neq 0$. We
find that $\xi _{y}(\rho \to 0)$ does depend on $\phi $ (the same can be
shown for $\xi _{z}$). This means that the vector $\xi ^{\mu }$ is {\em
discontinuous} at P.\footnote{
The indefiniteness of the RW self force could be intuitively understood, by
realizing that the RW gauge condition ``distracts'', to some amount, the
presumed isotropic structure of the divergent local piece of the metric
perturbation, by artificially signifying the $\theta $ direction. (This
isotropic structure is best accounted for within the harmonic gauge.)}

As the gauge displacement vector $\xi ^{\mu }$ does not admit a definite
value at the particle's location, Eq.\ (\ref{gg10}) cannot be used, as it
stands, for constructing the self force in the RW gauge. Following the
discussion at the end of Sec.\ \ref{secII}, we arrive at the conclusion that
{\em in the case of circular motion, the ``RW self force'' is ill defined}
(unless one further extends the regularization procedure---e.g., by
introducing an average over solid angle; see the discussion in Sec.\ \ref
{secII}).

We conclude this discussion with two remarks: First, though the
discontinuity of $\xi ^{\mu }$ was explicitly demonstrated here for circular
orbits, this conclusion should also apply to generic non-radial,
non-circular, orbits (for radial orbits, however, it was demonstrated above
that $\xi ^{\mu }$ is continuous).

Second, the above construction shows that for a suitable choice of the free
parameters (namely $c=\beta _{n}=0$) the component $\xi _{y}$ is {\em bounded
} at P. The same holds for $\xi _{z}$. This implies that $\xi _{\theta }$
and $\xi _{\varphi }$ are bounded (though discontinuous) at the particle's
location. It still remains to be checked, however, whether $\xi _{t}$ and $
\xi _{r}$ are bounded or not.

\subsection{Radiation gauge}

Finally, we examine the transformation of the self force to the so-called
``radiation gauge''. (We recall that, so far, the mode decomposition of
metric perturbations in Kerr spacetime has been formulated primarily within
the radiation gauge \cite{Chrzanowski}.) We shall consider here the simplest
possible case: a static particle in flat spacetime. As we shall shortly see,
even in this trivial case, the gauge transformation from the harmonic to the
radiation gauge is pathological, and the metric perturbation (and hence the
self force) is ill defined.

We shall specifically consider the {\it outgoing} radiation gauge (similar
results are obtained when considering the ingoing radiation gauge). We use
standard flat-space spherical coordinates $t,r,\theta ,\varphi $, and assume
that the static particle is located off the origin of the spherical
coordinates, i.e. at some $r=r_{0}>0$. Also, without loss of generality, we
locate the particle at the polar axis, $\theta =0$. The outgoing null vector
field takes the form $l^{\alpha }=[1,1,0,0]$. The metric perturbation in the
radiation gauge, $h_{\alpha \beta }^{{\rm (R)}}$, is defined by the
requirement\footnote{
In the case of a pure vacuum perturbation over a Kerr background, the
additional condition $h^{\rm (R)}\equiv g^{\alpha \beta }h_{\alpha \beta }^{
{\rm (R)}}=0$ can be imposed in a consistent manner, as done by Chrzanowski in
\cite{Chrzanowski}. Here we consider the perturbation in a region
surrounding a point source, and it is unclear to us whether the additional
condition $h^{\rm (R)}=0$ will be consistent with the gauge condition (\ref
{gg200}). We shall therefore not make any use of this extra condition here.}
\begin{equation}
h_{\alpha \beta }^{{\rm (R)}}l^{\beta }=0.  \label{gg200}
\end{equation}

Consider now the \thinspace $t$-component of Eq.\ (\ref{gg200}), which reads
\[
h_{tt}^{{\rm (R)}}+h_{tr}^{{\rm (R)}}=0\,.
\]
With the gauge transformation equation $h_{\alpha \beta }^{{\rm (R)}
}=h_{\alpha \beta }^{{\rm (H)}}+\xi _{\alpha ;\beta }+\xi _{\beta ;\alpha }$,
this becomes
\begin{equation}
\xi _{t,r}+\xi _{r,t}+2\xi _{t,t}=-h_{tt}^{{\rm (H)}}-h_{tr}^{{\rm (H)}}\,.
\label{gg201}
\end{equation}
Motivated by the staticity of the problem, we shall only consider
$t$-independent solutions, so $\xi _{r,t}$ and $\xi _{t,t}$ may be dropped.
Also, in the harmonic gauge we have
\[
h_{tt}^{{\rm (H)}}=2m/\epsilon,\quad\quad h_{tr}^{{\rm (H)}}=0,
\]
where $\epsilon $ denotes the spatial distance to the particle's location
(this may be easily obtained by transforming the well known Coulomb-like
Cartesian solution to spherical coordinates). Equation (\ref{gg201}) now
becomes
\begin{equation}
\xi _{t,r}=-2m/\epsilon \,.  \label{gg201a}
\end{equation}
At this point we introduce standard Cartesian coordinates $t,x,y,z\,$, such
that the particle is located at the origin ($x=y=z=0$), and the $z$
direction coincides with the radial direction at the particle's location
(namely, $x=r\sin \theta \cos \varphi $, $y=r\sin \theta \sin \varphi $, and
$z=r\cos \theta -r_{0}$). At the leading order in $\epsilon $, we may
replace $\partial _{r}$ by the Cartesian derivative operator $\partial
_{z}\, $. Equation (\ref{gg201a}) then becomes
\begin{equation}
\xi _{t,z}\cong -2m(z^{2}+\rho ^{2})^{-1/2}\,,  \label{gg202}
\end{equation}
where $\rho ^{2}\equiv x^{2}+y^{2}$. Eq.\ (\ref{gg202}) can now be easily
integrated with respect to $z$ (with $x,y$ held fixed). We obtain
\begin{equation}
\xi _{t}\cong -2m\,\log \left( \frac{z}{\rho }+\sqrt{1+z^{2}/\rho ^{2}}
\right) +R(x,y)\,,
\end{equation}
where $R(x,y)$ is an arbitrary function. This is the most general
($t$-independent) solution for $\xi _{t}$.

Consider next the asymptotic form of $\xi _{t}$ as we go to the limit $
x,y\to 0$ with fixed $z\ne 0$. One finds
\begin{equation}
\xi _{t}(\rho \to 0)\cong \left\{
\begin{array}{ll}
+2m\ln (\rho /2z)+R(x,y), & z>0, \\
-2m\ln (\rho /2|z|)+R(x,y), & z<0.
\end{array}
\right.  \label{gg204}
\end{equation}
By a suitable choice of the function $R(x,y)$ one may, at best, eliminate
the divergence along one of the rays $z<0$ or $z>0$ (by choosing $R\simeq
\pm 2m\ln \rho $, respectively), but not along both rays simultaneously. We
thus arrive at the conclusion that $\xi _{t}$ unavoidably diverges
logarithmically (at least) on approaching the axis $\rho =0$, along either
the $z<0$ ray or the $z>0$ ray (or both). Constructing now the $tx$ and $ty$
components of the radiation-gauge metric perturbation, we find $h_{tx}^{{\rm
(R)}}=\xi _{t,x}\propto x/\rho ^{2}$, and a similar expression for $h_{ty}^{
{\rm (R)}}$, as $\rho \to 0$ (at either $z<0$ or $z>0$). Namely, the metric
perturbation inevitably diverges at least along half the axis $\rho =0$.

It thus turns out that in the radiation gauge, the perturbation associated
with a pointlike particle is represented by a string-like one-dimensional
singularity. In particular, the radiation-gauge metric perturbation cannot
be well defined in a complete neighborhood of the particle. (Compare with
the harmonic or RW gauges, where the singularity is confined to the
particle's location and the metric perturbation is well defined everywhere
in the particle's neighborhood.) This pathological behavior---manifested
already in the elementary case of a static particle in flat space---serves
to demonstrate the pathological nature of the radiation gauge in the
presence of point sources. As the radiation gauge seems inappropriate for
representing the metric perturbation in the particle's neighborhood, it
becomes rather meaningless to consider the self force acting on the particle
in that gauge.

Finally we note that although the indefiniteness of the radiation-reaction
self force was demonstrated here only for a static particle in flat space, the
same indefiniteness should also occur generically for all types of orbits in
Schwarzschild or Kerr spacetimes.


\section{Summary and discussion}

\label{secVI}

The main results of this manuscript are contained in Eqs.\ (\ref{gg10}),
(\ref{gg45}), and (\ref{eqVI170}). Eq.\ (\ref{gg10}) describes the gauge
transformation of the gravitational self force, given the gauge displacement
vector $\xi _{\mu }$. Eq.\ (\ref{gg45}) describes, in a schematic manner,
the extension of the MSTQW formulation for the gravitational self force to
an arbitrary gauge `G': It implies that the ``correct'' singular piece to be
removed from the bare force in the G-gauge [the one derived directly from
the G-gauge metric perturbation through Eq.\ (\ref{gg34})] is always to be
calculated {\em in the harmonic gauge}, as described in the original
analysis by MSTQW. By applying these results to our mode-sum regularization
method (which was previously formulated only within the harmonic gauge \cite
{paperI}) we finally obtained Eq.\ (\ref{eqVI170}), which describes a
practical mode-sum prescription for construction of the gravitational self
force in any gauge `G' (provided that the self force has a regular, finite
value in that gauge). We stress again that, since the gravitational self
force is a gauge-dependent notion, expressions like Eq.\ (\ref{gg45})
or Eq.\ (\ref{eqVI170}) for the self force will be meaningful only when
accompanied by the full information about the gauge to which they correspond.
(Alternatively, one can specify the metric perturbation $h^{\rm (G)}$ itself,
which of course contains the full information about the gauge.)

The implementation of Eq.\ (\ref{eqVI170}) for calculating the G-gauge self
force involves two distinct parts: (i) calculation of the bare modes of the
force in the G-gauge [through Eq.\ (\ref{gg34})]; and (ii) derivation of
the four vectorial regularization parameters $A^{\alpha }$, $B^{\alpha }$, $
C^{\alpha }$, and $D^{\alpha }$. Our discussion concerning the gauge
transformation of the self force led us to conclude that the values of these
regularization parameters do not depend on the gauge in which one calculates
the self force: These parameters are always to be calculated in the harmonic
gauge (using the analytic technique described in Ref.\ \cite{paperI}). This
``gauge invariance'' property of the regularization parameters is
demonstrated by the recent analysis by Lousto \cite{Lousto,BL}, who
calculated (numerically) the values of $A^{\alpha }$, $B^{\alpha }$, and $
C^{\alpha }$ in the RW gauge, for a radial orbit on a Schwarzschild
background. These numerical values appear to be in perfect agreement
with the harmonic-gauge values derived analytically in Ref.\ \cite{paperI}
(in the case studied so far, of the self force at a turning point of a
radial geodesic). Also, the (zero) value obtained for the parameter $
D^{\alpha }$ in the harmonic gauge \cite{paperI} agrees with Lousto's result
for $D^{\alpha }$ in the RW-gauge (which was based on a proposed
zeta-function regularization procedure \cite{Lousto}).

The prescription (\ref{eqVI170}), as well as Eq.\ (\ref{gg45}), is only
applicable when the self force admits a definite finite value in the
G-gauge. Whether or not this is the case for a given gauge ``G'', can be
decided with the help of Eq.\ (\ref{gg10}): The analysis by MSTQW implies
that the self force will always have a regular finite value in the harmonic
gauge (and it also tells us how to derive this value). Therefore, the
G-gauge self force would be well defined, in our approach, only if the
transformation from the harmonic gauge to the G-gauge would yield---through
Eq.\ (\ref{gg10})---a regular finite value for the force difference $\delta
F_{{\rm self}}^{\alpha }$. It is only in this case that we are able to use
Eq.\ (\ref{eqVI170}) for calculating the G-gauge self force. Otherwise
(namely, if $\delta F_{{\rm self}}^{\alpha }$ diverges or is indefinite),
Eq.\ (\ref{eqVI170}) appears to be useless.

As an example, in Sec.\ \ref{secV} we explored the transformation from the
harmonic gauge to the Regge-Wheeler gauge. We found that the RW self force
is well defined as long as strictly radial trajectories are considered. For
such trajectories, Eq.\ (\ref{eqVI170}) then provides a useful prescription
for computing the RW self force. However, this seems not to be the case for
more general orbits, as we demonstrated by considering a circular orbit:
Here, the transformation from the harmonic gauge yielded an indefinite value
for the RW self force. The situation is even worse in the radiation gauge,
where $\delta F_{{\rm self}}^{\alpha{\rm (H\rightarrow R)}}$ is found
to be not only discontinuous but also unbounded, and presumably for {\em all}
types of orbits.

How could one interpret a situation where $\delta F_{{\rm self}}^{\alpha }$
diverges (or is indefinite)? In some occasions, such a result may be
attributed to a severe pathology of the gauge. This seems to be the case in
the radiation gauge, as implied by the fact that in this gauge the metric
perturbation diverges not only at the particle's location, but also along an
(ingoing or outgoing) radial ray emerging from the particle (see Sec.\ \ref
{secV}). However, the situation seems to be different in the RW gauge, in
which the metric perturbation is well-defined in the neighborhood of the
particle (though of course not at the particle itself), like in the harmonic
gauge. In this case we have seen that, for non-radial orbits, $\delta F_{
{\rm self}}^{\alpha{\rm (H\rightarrow RW)}}$ (and hence also $F_{{\rm
self}}^{\alpha{\rm (RW)}}$ itself) is ill defined. This originates from
the fact that certain components of $\xi ^{\mu }$---e.g. $\xi ^{\theta }$
or $\xi ^{\varphi }$---admit a direction-dependent limit (as demonstrated
by the dependence of e.g. $\xi _{y}$ on $\phi $; cf. Sec.\ \ref{secV}).

This situation---a direction-dependent expression for the self force in
certain gauges---motivates one to consider a simple generalization of the
standard MSTQW regularization procedure, by averaging over all spatial
directions. Namely, one can evaluate the limit of the right-hand side of
Eq.\
(\ref{gg45}) (or, similarly, the limit of the displacement vector $\xi ^{\mu
}$) along fixed spatial null geodesics emanating from the particle, and then
average over the solid angle (in the particle's rest frame). This would
clearly be a generalization of the MSTQW procedure, because whenever the
coincidence limit is well defined, the average over solid angle will be
well-defined too, and will yield the same result. One still needs to
investigate how this averaging over directions is to be implemented within
the context of the mode-sum regularization.

The above generalized regularization procedure will yield a definite self
force in a wide class of gauges (though not in all gauges; Obviously one can
construct a displacement vector $\xi ^{\mu }$ which does not even have a
directional limit, in which case the generalized regularization procedure
will fail to yield a definite self force). The analysis in Sec.\ \ref{secV}
suggests that for circular orbits the displacement vector $\xi ^{\mu }$ from
the harmonic to the RW gauge may have a well-defined directional limit, and
hence the RW self force may be well defined within this generalized
prescription. Recall, however, that the above analysis does not completely
guarantee this regularity of the (generalized) RW self force, because so far
we have only analyzed the tangential components $\xi _{y}$ and $\xi _{z}$
(which yield $\xi ^{\theta }$ and $\xi ^{\varphi }$), but not $\xi _{t}$ and
$\xi _{r}$. Also, our analysis was restricted to the surface $r=r_{0},t=0$,
i.e. to directional limits through tangential directions.

There seems to be another procedure that would allow one to use the metric
perturbations in e.g. the RW or radiation gauges for useful self-force
calculations (without resorting to the above generalized regularization
procedure). We shall now briefly outline here a preliminary version of this
procedure.
(We note that a similar approach has been proposed by Mino\cite{Mino}.)
Suppose that the metric perturbation $h^{{\rm (G)}}$ is known (e.g.
in the form of mode decomposition), where ``G'' refers to either the RW or
radiation gauges. If we knew how to convert $h^{{\rm (G)}}$ to the harmonic
gauge, it would be straightforward to construct the self force from it,
through Eqs.\ (\ref{gg31}) or (\ref{eqVI160}). However, performing the
transformation G$\rightarrow $H requires one to solve a system of partial
differential equations for $\xi ^{\mu }$, and unfortunately we do not know
the exact solution of this system. Nevertheless, it appears possible to
construct an approximate, leading-order, solution of this system, for both
the RW and the radiation gauges. This was demonstrated in Sec.\ \ref{secV}
(for both gauges) in a few simple cases, and it appears likely that the
leading-order solution can be generalized to a generic orbit. Let us denote
this leading-order solution by $\hat{\xi}^{\mu }$. In principle one can then
use $\hat{\xi}^{\mu }$ to transform the metric perturbations from the
original gauge G to an ``approximate harmonic'' gauge, which we denote
\^{H}. Presumably, in the gauge \^{H} the self force will be well defined,
since the metric perturbations in the harmonic and \^{H} gauges share the
same leading-order asymptotic behavior. After decomposing $\hat{\xi}^{\mu }$
into $l$-modes, one can use the mode-sum regularization method to
calculate the \^{H}-gauge self force:
Applying Eq.\ (\ref{gg40}) for each of the single $l$-modes, with ``H''
and ``G'' replaced, correspondingly, by ``G'' and ``\^{H}'', we first
get
\[
F^{\alpha l{\rm (\hat H)}}_{\rm bare}=F^{\alpha l{\rm (G)}}_{\rm bare}
+\delta F^{\alpha l{\rm (G\to \hat{H})}}_{\rm self},
\]
where $\delta F^{\alpha l{\rm (G\to \hat H)}}_{\rm self}$ is to be obtained
from Eq.\ (\ref{gg10}) by replacing $\xi ^{\lambda }$ by the $l$-mode of
$\hat{\xi}^{\mu }$.
Then, writing Eq.\ (\ref{eqVI170}) for the \^{H}-gauge (i.e., with all
``G'' replaced by ``\^{H}'') and substituting the above expression
for $F^{\alpha l{\rm (\hat H)}}_{\rm bare}$,
one obtains
\begin{equation}
F^{\alpha{\rm (\hat{H})}}_{\rm self}=
\sum_{l=0}^{\infty }\left[\left(F^{\alpha l{\rm (G)}}_{\rm bare}
+\delta F^{\alpha l{\rm (G\to \hat H)}}_{\rm self}\right)
-A^{\alpha }L-B^{\alpha }-C^{\alpha }/L\right] -D^{\alpha },
\end{equation}
which provides a prescription for calculating the \^{H}-gauge self
force through the modes of the bare force in the G gauge.
We hope to further develop and implement this method elsewhere.

\section*{Acknowledgements}

It is our pleasure to thank Eric Poisson, Carlos Lousto, and Yasushi Mino
for commenting on an early version of this manuscript, and for valuable
discussions. This research was supported in part by the United
States-Israel Binational Science Foundation.
L.B.\ was also supported by a Marie Curie Fellowship of
the European Community programme IHP-MCIF-99-1 under contract number
HPMF-CT-2000-00851.



\end{document}